\documentclass{ws-procs9x6}
\usepackage[T1]{fontenc}
\usepackage[latin1]{inputenc}
\usepackage{prettyref}
\usepackage{amsmath}
\usepackage{graphicx}
\usepackage{amssymb}
\usepackage{bm}
\usepackage{color}
\makeatletter
\usepackage{ae,aecompl}
\makeatother
\begin{document}

\title{Neutrino Processes in Strong Magnetic Fields}

\author{Huaiyu Duan}

\address{Department of Physics\\
University of California, San Diego\\
La Jolla, CA 92093, USA\\
E-mail: hduan@ucsd.edu}

\author{Yong-Zhong Qian}

\address{School of Physics and Astronomy\\
University of Minnesota\\
Minneapolis, MN 55455, USA\\
E-mail: qian@physics.umn.edu}

\maketitle

\abstracts{
The processes $\nu_{e}+n\rightleftharpoons e^{-}+p$ and 
$\bar{\nu}_{e}+p\rightleftharpoons e^{+}+n$
provide the dominant mechanisms for heating and cooling the material
below the stalled shock in a core-collapse supernova. We summarize 
the major effects of strong magnetic fields on the rates of the 
above reactions and illustrate these effects with a simple supernova 
model. Due to parity violation of weak interaction
the heating rates are asymmetric even for a uniform magnetic field.
The cooling rates are also asymmetric for nonuniform fields.
The most dramatic effect of strong magnetic fields of $\sim 10^{16}$ G
is suppression of the cooling rates by changing the equations of
state through the phase space of $e^{-}$ and $e^{+}$.}

\section{Introduction}

The neutrino processes\begin{eqnarray}
\nu_{e}+n & \rightleftharpoons & e^{-}+p,\label{eq:nun}\\
\bar{\nu}_{e}+p & \rightleftharpoons & e^{+}+n\label{eq:nup}\end{eqnarray}
play important roles in core-collapse supernovae. After the shock
is stalled, neutrinos emitted from the protoneutron star exchange
energy with the material below the shock mainly through these processes. 
The forward processes in Eqs.~\prettyref{eq:nun} and \prettyref{eq:nup}
heat the material through the absorption of $\nu_{e}$ and $\bar{\nu}_{e}$,
while the reverse processes cool the material by emitting them. In
the neutrino-driven supernova mechanism [\refcite{Bethe:1985}], the competition
between heating and cooling of the material by these processes is
expected to result in net energy gain for the stalled shock, which is
then revived to make a successful supernova
explosion. Unfortunately, the current consensus is that this
mechanism does not work in spherically symmetric models
[\refcite{Rampp:2000,Liebendorfer:2001}]. 
On the other hand, strong magnetic fields may be generated
during the formation of protoneutron stars and in turn affect supernova
dynamics. Observations have shown that some neutron stars possess
magnetic fields as strong as $\sim10^{15}\ \textrm{G}$ 
[\refcite{Kouveliotou:1999}--\refcite{Ibrahim:2003}].
Although it is not clear how strong magnetic fields in supernovae could be,
some calculations indicate that fields of
$10^{15}$--$10^{16}\ \textrm{G}$
are not impossible [\refcite{Akiyama:2003}].
While strong magnetic fields can affect supernova dynamics in many
possible ways, here we consider their effects on the neutrino
processes. Because the explosion energy is much smaller than the
gravitational binding energy of the protoneutron star and nearly all of 
the latter is released in neutrinos, it
is natural to expect that a small change in the neutrino physics input
may have a large impact on the supernova mechanism. 

The effects of strong magnetic fields on neutrino processes have been
studied in various approximations 
[\refcite{Roulet:1998sw}--\refcite{Bhattacharya:2002qf}].
In our recent work, we have calculated the effects of magnetic fields
on the four processes in Eqs.~\prettyref{eq:nun} and \prettyref{eq:nup}
to the 0th [\refcite{Duan:2004nc}] and 1st order [\refcite{dq}]
in $E_{\nu}/m_{N}$, where $E_{\nu}$ is the neutrino energy and 
$m_{N}$ is the nucleon mass. Here we summarize our results and list
some issues that remain to be addressed.

\section{Neutrino Processes in Strong Magnetic Fields}

\subsection{General Effects of Magnetic Fields}

An obvious effect of the magnetic field is polarization of the
spin of a nonrelativistic nucleon. When this effect is
small, the polarization of the nucleon spin may be written as
\begin{equation}
\chi\simeq\frac{\mu B}{T}=3.15\times10^{-2}
\left(\frac{\mu}{\mu_{N}}\right)\left(\frac{B}{10^{16}\ \textrm{G}}\right)
\left(\frac{\textrm{MeV}}{T}\right),
\label{eq:polarization}
\end{equation}
where $\mu$ is the nucleon magnetic moment, $\mu_{N}=e/2m_{p}$ 
is the nuclear magneton, $B$ is the magnetic field strength, and $T$ is
the gas temperature. Due to parity violation of weak interaction
polarization of the nucleon spin introduces a dependence on the angle 
$\Theta_{\nu}$ between the directions of the neutrino momentum and 
the magnetic field for the cross sections of the forward processes
in Eqs.~\prettyref{eq:nun} and \prettyref{eq:nup} (see Sec. 2.2)
and for the differential volume reaction rates 
of the reverse processes (see Sec. 2.3).

In addition, assuming a magnetic field in the positive $z$-direction,
the motion of a proton in the $xy$-plane is quantized into Landau levels 
(see e.g.~Ref.~[\refcite{Landau:1977}])
with kinetic energies
\begin{equation}
E_{p}(n_{p},k_{pz})=\frac{k_{pz}^{2}}{2m_{p}}+\left(n_{p}+
\frac{1}{2}\right)\frac{eB}{m_{p}},\qquad n_{p}=0,1,2,\cdots,
\label{eq:energy-nonrel}
\end{equation}
where $n_{p}$ is the quantum number of the proton Landau level and
$k_{pz}$ is the $z$-component of the proton momentum. 
We are interested in gas temperatures of $T\gtrsim1\ \mathrm{MeV}$ and 
magnetic fields of $B\sim10^{16}\ \mathrm{G}$.
As $eB/m_{p}=63(B/10^{16}\ \mathrm{G})\ \mathrm{keV}$, for such
conditions a proton is able to occupy Landau levels with $n_{p}\gg1$
and can be considered as classical.

For the conditions of interest here, $e^{-}$ and $e^{+}$ are relativistic.
Their Landau levels have energies
\begin{equation}
E_{e}(n_{e},k_{ez})=\sqrt{m_{e}^{2}+k_{ez}^{2}+2n_{e}eB},
\label{eq:energy-rel}
\end{equation}
where symbols are defined similarly to those for the proton.
The above equation has taken spin into account. Note that the $e^{-}$ 
or $e^{+}$ in the ground Landau level ($n_{e}=0$) has only one spin state.
This introduces an additional dependence on $\Theta_{\nu}$ (independent of
polarization of the nucleon spin) for the cross sections of the forward 
processes in Eqs.~\prettyref{eq:nun} and \prettyref{eq:nup} (see Sec. 2.2)
and for the differential volume reaction rates 
of the reverse processes (see Sec. 2.3).
The effects of Landau levels are more prominent for $e^{-}$ and $e^{+}$ 
than for nucleons. This can be seen from the quantum number for the highest
Landau level occupied by $e^{-}$ or $e^{+}$ with energy $E_{e}$,
\begin{equation}
(n_{e})_{\mathrm{max}}=\left[\frac{E_{e}^{2}-m_{e}^{2}}
{2eB}\right]_{\mathrm{int}}=\left[8.45\times10^{-3}
\left(\frac{E_{e}^{2}-m_{e}^{2}}{\mathrm{MeV}^{2}}\right)
\left(\frac{10^{16}\ \mathrm{G}}{B}\right)\right]_{\mathrm{int}}.
\end{equation}
To account for the effects of Landau levels, part of the
integration over the phase space of $e^{-}$ or $e^{+}$ is changed
to a summation over possible Landau levels, i.e.
\begin{equation}
2\int\frac{\textrm{d}^{3}k_e}{(2\pi)^{3}}\rightarrow
\frac{eB}{2\pi}\sum_{n_{e}=0}^{(n_{e})_{\mathrm{max}}}g_{n_{e}}
\int\frac{\textrm{d}k_{ez}}{2\pi},
\label{eq:phase-space}
\end{equation}
where $g_{n_{e}}$ is the number of spin states for the $n_{e}$th
Landau level ($g_{n_{e}}=1$ for $n_{e}=0$ and 2 for $n_{e}>0$).

\subsection{Heating Processes}

To the 0th order in $E_{\nu}/m_{N}$, the cross sections of the forward
processes in Eqs.~\prettyref{eq:nun} and \prettyref{eq:nup} are
found to be [\refcite{Duan:2004nc}]
\begin{eqnarray}
\sigma_{\nu N}^{(0,B)} &= &
\sigma_{B,1}\left[1+2\chi\frac{(f\pm g)g}{f^{2}+3g^{2}}\cos\Theta_{\nu}\right]
\nonumber \\
& & +\sigma_{B,2}\left[\frac{f^{2}-g^{2}}{f^{2}+3g^{2}}
\cos\Theta_{\nu}+2\chi\frac{(f\mp g)g}{f^{2}+3g^{2}}\right],
\label{eq:sigma-nuN-b}
\end{eqnarray}
where the energy-dependent factors $\sigma_{B,1}$ and $\sigma_{B,2}$
are defined as
\begin{eqnarray}
\sigma_{B,1} & = & \frac{G_{\mathrm{F}}^{2}\cos^{2}\theta_{\mathrm{C}}}{2\pi}
(f^{2}+3g^{2})eB\sum_{n_{e}=0}^{(n_{e})_{\mathrm{max}}}
\frac{g_{n_{e}}E_{e}}{\sqrt{E_{e}^{2}-m_{e}^{2}-2n_{e}eB}},\\
\sigma_{B,2} & = & \frac{G_{\mathrm{F}}^{2}\cos^{2}\theta_{\mathrm{C}}}{2\pi}
(f^{2}+3g^{2})eB\frac{E_{e}}{\sqrt{E_{e}^{2}-m_{e}^{2}}},
\end{eqnarray}
with $E_{e}=E_\nu\pm\Delta$ and $\Delta$ being the neutron-proton mass
difference. In the above equations,
the upper sign is for $\nu_e$ absorption on
$n$ and the lower sign for $\bar\nu_e$ absorption on $p$.

\begin{figure}
\begin{center}\includegraphics[%
  clip,
  width=1.0\textwidth,
  keepaspectratio]{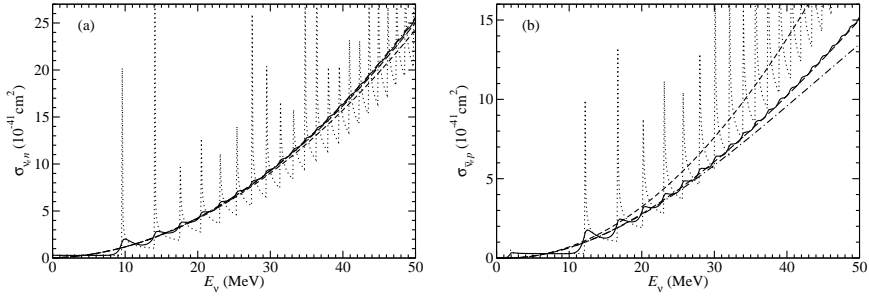}  
\end{center}

\caption{\label{fig:sigma-nuN}The cross sections of 
$\nu_{e}+n\rightarrow e^{-}+p$
(a) and $\bar{\nu}_{e}+p\rightarrow e^{+}+n$ (b) as functions of
neutrino energy $E_{\nu}$. The angle $\Theta_{\nu}$ between the directions
of the neutrino momentum and the magnetic field is taken to be
0. The dotted and solid curves are the cross sections to the 0th and
1st order in $E_{\nu}/m_{N}$, respectively. Both assume a magnetic
field of $B=10^{16}\ \textrm{G}$. In addition, the solid curves assume
a temperature $T=2\ \textrm{MeV}$ for the nucleon gas. The short-dashed 
and dot-dashed curves are the cross sections to the 0th and 1st order
in $E_{\nu}/m_{N}$, respectively, but for $B=0$. The long-dashed
curve in (b) is for $B=0$ and includes some 
corrections beyond the 1st order. The differences between
the short-dashed, dot-dashed, and long-dashed curves in (b) at the
high-energy end are mostly due to the combined effects of nucleon
recoil and weak magnetism.}
\end{figure}

Comparing $\sigma_{B,1}$ with the well-known expression for the
0th-order cross section in the absence of magnetic fields,
\begin{equation}
\sigma_{\nu N}^{(0)}=
\frac{G_{\mathrm{F}}^{2}\cos^{2}\theta_{\mathrm{C}}}{\pi}
(f^{2}+3g^{2})k_{e}E_{e},
\end{equation}
one can see that the only difference is the change in phase space
{[}see Eq.~\prettyref{eq:phase-space}{]}. To illustrate the effects
of strong magnetic fields, we plot the cross sections of neutrino
absorption on nucleons for $B=10^{16}$ G and 0
in Fig.~\ref{fig:sigma-nuN} (see Ref.~[\refcite{dq}] for more details).
An immediate observation is that the cross sections are
enhanced (dotted curves in Fig.~\ref{fig:sigma-nuN})
if the energy of the outgoing $e^{-}$ or $e^{+}$ satisfies the condition
\begin{equation}
E_{e}=\sqrt{m_{e}^{2}+2n_{e}eB}
\label{eq:peak-cond}
\end{equation}
for $n_{e}>0$. This is because a new Landau
level opens up when Eq.~\prettyref{eq:peak-cond} is satisfied. Just
as discrete energy levels of atoms lead to absorption lines in the
light spectra, ideally the presence of strong magnetic fields would
produce sharp dips in the neutrino energy spectra where 
Eq.~\prettyref{eq:peak-cond} holds.
However, the nucleons absorbing neutrinos have thermal
motion, which will smear out these sharp dips. We have included
the thermal motion of nucleons and calculated the cross sections
to the 1st order in $E_{\nu}/m_{N}$ [\refcite{dq}]. Even for magnetic fields
as strong as $10^{16}\ \textrm{G}$, the thermal motion of nucleons
with $T\sim2\ \textrm{MeV}$ is enough to smooth out almost
all the spikes in $\sigma_{\nu N}(E_{\nu})$ (see solid curves in 
Fig.~\ref{fig:sigma-nuN}). The effect of the magnetic field is further
diminished by averaging the cross sections over neutrino energy spectra.
A magnetic field of $10^{16}\ \textrm{G}$ causes changes of
only a few percent to the average cross sections.

\begin{figure}
\begin{center}\includegraphics[%
  clip,
  width=0.50\textwidth,
  keepaspectratio]{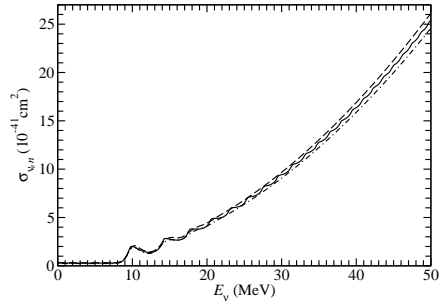}
\end{center}

\caption{\label{fig:sigma-nun-ang}The cross section of 
$\nu_{e}+n\rightarrow e^{-}+p$
to the 1st order in $E_{\nu}/m_{N}$ for $B=10^{16}\ \textrm{G}$ and
$T=2\ \textrm{MeV}$. The angle $\Theta_{\nu}$ between
the directions of $\nu_{e}$ and the magnetic field is taken to be $0$
(dot-dashed curve), $\pi/2$ (solid curve), and $\pi$ (dashed curve),
respectively.}
\end{figure}

The term proportional to $\sigma_{B,1}$ in Eq.~\prettyref{eq:sigma-nuN-b}
depends on the direction of the incoming neutrino with respect to the
magnetic field. This is due to parity violation of weak interaction. 
In fact, as long as the target nucleon has
a polarization $\chi$, the lowest-order expression of the cross section
in the absence of magnetic fields can be written as
\begin{equation}
\sigma_{\nu N}^{(0)}(\chi)=\sigma_{\nu N}^{(0)}\left[1+2\chi
\frac{(f\pm g)g}{f^{2}+3g^{2}}\cos\Theta_{\nu}\right],
\label{eq:sigma-angle}
\end{equation}
which has exactly the same angular dependence as the $\sigma_{B,1}$ term 
in Eq.~\prettyref{eq:sigma-nuN-b}. The appearance of the
term proportional to $\sigma_{B,2}$ in this equation
is due to the fact that there is only one spin state for the ground
Landau level of $e^{-}$ or $e^{+}$. This term has an angular dependence 
even if the nucleon polarization $\chi=0$. Nevertheless, this
dependence is again due to parity violation as the $e^{-}$ or $e^{+}$
in the ground Landau level is polarized. For crude estimates of
the angular dependence, we use Eq.~\prettyref{eq:sigma-angle}.
As the nucleon form factors 
$f=1$ and $g=1.26$ are close in numerical value, 
$\sigma_{\bar{\nu}_{e}p}^{(0)}(\chi_p)$ has little dependence on 
$\Theta_{\nu}$. On the other hand, the $\Theta_{\nu}$-dependent term 
for $\sigma_{\nu_{e}n}^{(0)}(\chi_n)$ is
$\sim\chi_{n}\cos\Theta_{\nu}$. The angular dependence
of $\sigma_{\nu_{e}n}$ for a strong magnetic field of $B=10^{16}$ G
and a gas temperature of $T=2$ MeV is shown in Fig.~\ref{fig:sigma-nun-ang}.

\subsection{Cooling Processes}

Because $e^{-}$ and $e^{+}$ do not have definite velocities 
[\refcite{Landau:1977}],
we define a volume reaction rate $\Gamma_{eN}$, which gives the rate
of e.g., $e^{+}$ capture per neutron when multiplied by the $e^{+}$
number density $\mathfrak{n}_{e^{+}}$. The
differential volume reaction rates to the 0th order in $E_{\nu}/m_{N}$
are found to be [\refcite{Duan:2004nc}]
\begin{eqnarray}
\frac{\textrm{d}\Gamma_{eN}^{(0,B)}}{\mathrm{d}\cos\Theta_{\nu}} 
& = & \frac{\Gamma_{eN}^{(0)}}{2}\left[1+2\chi\frac{(f\pm g)g}
{f^{2}+3g^{2}}\cos\Theta_{\nu}\right]
\nonumber \\
& & +\delta_{n_{e},0}\frac{\Gamma_{eN}^{(0)}}{2}
\left[\frac{f^{2}-g^{2}}{f^{2}+3g^{2}}\cos\Theta_{\nu}+
2\chi\frac{(f\mp g)g}{f^{2}+3g^{2}}\right],
\label{eq:dGamma-eN}
\end{eqnarray}
where 
\begin{equation}
\Gamma_{eN}^{(0)}=
\frac{G_{\mathrm{F}}^{2}\cos^{2}\theta_{\mathrm{C}}}{2\pi}
(f^{2}+3g^{2})E_{\nu}^{2}
\end{equation}
with $E_\nu=E_e\pm\Delta$
is the volume reaction rate in the absence of magnetic fields. In the
above equations, the upper sign is for $e^{+}$ capture
on $n$ and the lower sign for $e^{-}$ capture on $p$. 
The angular dependence for neutrino emission in Eq.~\prettyref{eq:dGamma-eN} 
is the same as that for neutrino absorption in Eq.~\prettyref{eq:sigma-nuN-b}
and is due to parity violation of weak interaction as explained in Sec. 2.2.
However, the volume reaction rates for the cooling processes obtained by 
integrating the differential rates in Eq.~\prettyref{eq:dGamma-eN} over 
$\Theta_{\nu}$ are isotropic for a uniform magnetic field. This is in 
contrast to the cross sections in Eq.~\prettyref{eq:sigma-nuN-b} for 
the heating processes.

\begin{figure}
\begin{center}\includegraphics[%
  width=1.0\textwidth]{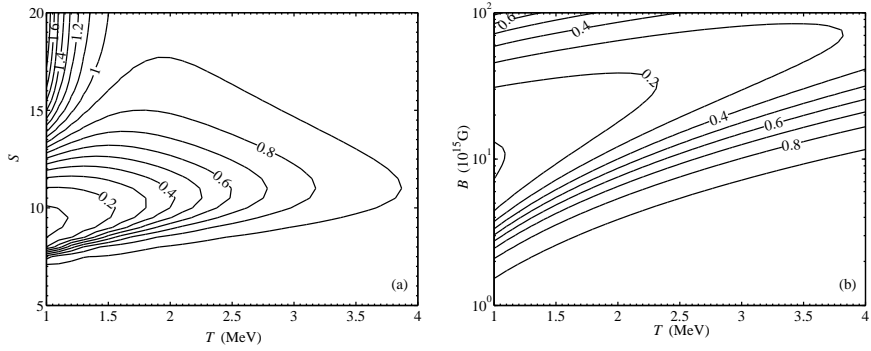}\end{center}

\caption{\label{fig:cooling-suppression}Contours of the ratio of the total
cooling rate with magnetic fields to that without.
A magnetic field of $10^{16}\ \textrm{G}$ is assumed for (a) and 
a total entropy per nucleon $S=10$ is assumed for (b).
A constant electron fraction $Y_e=0.5$ is assumed for both.}
\end{figure}

Like the cross sections for the heating processes, the volume reaction 
rates of the cooling
processes are not much affected even for magnetic fields as strong
as $10^{16}\ \mathrm{G}$. However, the cooling rates could be
severely suppressed by such strong magnetic fields due to changes
in the equations of state through the phase space of $e^{-}$ and $e^{+}$
[\refcite{Duan:2004nc}]. Given the electron fraction $Y_{e}$, the total
entropy per nucleon $S$, and the gas temperature $T$,
one can solve the equations of state
\begin{eqnarray}
\frac{\rho Y_{e}}{m_{N}} & = & \mathfrak{n}_{e^{-}}-\mathfrak{n}_{e^{+}},
\label{eq:eos1}\\
S & = & S_{N}+S_{\gamma}+S_{e^{-}}+S_{e^{+}},
\label{eq:eos2}
\end{eqnarray}
to obtain the baryon mass density $\rho$ and the electron degeneracy 
parameter $\eta_{e}$ for cases of strong and no magnetic fields.
The cooling rates (per nucleon) in each case
can then be calculated by integrating 
the volume reaction rates over the energy-differential number densities of
$e^{-}$ and $e^{+}$. 
The suppression of the total cooling rate by strong
magnetic fields is shown in Fig.~\ref{fig:cooling-suppression}.
The reason for this suppression is that compared with the case of no
magnetic fields, there are more low-energy $e^{-}$ and $e^{+}$ in
magnetic fields of $\sim 10^{16}\ \mathrm{G}$ as 
most of the $e^{-}$ and $e^{+}$ reside in the
ground Landau level ($n_{e}=0$). 

\section{Application to Core-Collapse Supernovae}

To illustrate the effects of strong magnetic fields on supernova dynamics,
we consider a simple supernova model. All neutrinos are assumed
to be emitted at the same radius $R_{\nu}=50\ \mathrm{km}$. The shock
is stalled at a radius $R_{\mathrm{s}}=200\ \mathrm{km}$. The electron
fraction and the total entropy per nucleon are taken to be
$Y_{e}=0.5$ and $S=10$, respectively, and held constant
between $R_{\nu}$ and $R_{\mathrm{s}}$. 
We adopt the temperature profile
\begin{equation}
T(r)=T(R_{\nu})\frac{R_{\nu}}{r}=4\left(\frac{50\ \mathrm{km}}{r}\right)\
\mathrm{MeV}.
\end{equation}
With the above assumptions, we calculate the total heating and cooling rates
as functions of radius $r$ for cases of strong and no magnetic fields
[the equations of state \prettyref{eq:eos1} and \prettyref{eq:eos2}
are solved to obtain $\rho$ and $\eta_e$ for calculating the
total cooling rate in both cases]. In the absence of magnetic fields, 
the total heating rate is found to be equal to the total cooling rate at
a gain radius $R_{\mathrm{g}}=137\ \mathrm{km}$,
above which heating dominates cooling. 
\vskip 0.1in
\begin{figure}
\begin{center}\includegraphics[%
  width=0.50\textwidth,
  keepaspectratio]{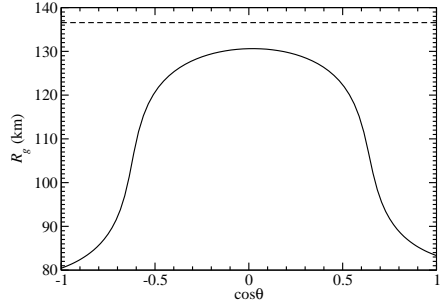}\end{center}

\caption{\label{fig:Rg-dipole}The gain radius $R_{\textrm{g}}$ 
as a function of $\cos\theta$
(solid curve) for a dipole magnetic field. Compared with the case
for $B=0$ (dashed curve), the gain radius is substantially
reduced at the north and south poles where the magnetic fields are
the strongest.}
\end{figure}

Consider a magnetic
field of dipole configuration in spherical coordinates $(r,\theta,\phi)$,
\begin{equation}
\mathbf{B}=B_{0}\left(\frac{R_{\nu}}{r}\right)^{3}
(2\cos\theta\mathbf{\hat{r}}+\sin\theta\mathbf{\hat{\bm\theta}}).
\end{equation}
We calculate $R_{\mathrm{g}}$ for $B_{0}=5\times10^{16}\ \mathrm{G}$ and 
plot it as a function of $\cos\theta$ in Fig.~\ref{fig:Rg-dipole}.
Due to suppression of the cooling rates by strong magnetic
fields, $R_{\mathrm{g}}$ becomes smaller in the presence of strong
magnetic fields compared with the case of no magnetic fields. 
As a result, there is more region of net heating below the stalled shock 
and the neutrino-driven supernova mechanism may work more efficiently. 
The cooling rates are most suppressed at north and south poles where 
the magnetic fields are the strongest. The stalled shock is likely to be 
revived earlier and more energetically in these directions. It is also worth
mentioning that there is a small difference in the gain radius between
the north and south poles (see Fig.~\ref{fig:Rg-dipole}).
This is due to the angular dependence of the heating processes discussed
in Sec. 2.2. This small asymmetry may result in a kick to the protoneutron
star that could explain the space velocities observed for pulsars.

\section{Open Issues}

Although we have demonstrated that strong magnetic
fields have important effects on the dynamics of core-collapse supernovae,
our results depend on how strong the magnetic fields in supernovae could 
be. This is the biggest open issue. Alternatively, our results can be
used to gauge whether magnetic fields would affect supernova dynamics
by changing the rates of neutrino processes. In this regard, we find
that magnetic fields weaker than $10^{15}\ \textrm{G}$ would have
negligible effects on the neutrino processes, while fields of 
$\sim 10^{16}\ \textrm{G}$ would dramatically
change supernova dynamics through neutrino physics.
Of course, magnetic fields weaker than $10^{15}\ \textrm{G}$ may already 
have important hydrodynamic effects in supernovae.
This is not considered here but should be investigated by future
studies. Another open issue is how to model supernova explosions
by including both hydrodynamic effects and changes in the neutrino
processes due to strong magnetic fields of $\sim 10^{16}\ \textrm{G}$.
The processes $\nu_{e}+n\rightarrow e^{-}+p$
and $\bar{\nu}_{e}+p\rightarrow e^{+}+n$ not only provide the
dominant mechanisms for heating the material below the stalled shock,
but also are the main opacity sources for determining the thermal
decoupling of $\nu_{e}$ and $\bar{\nu}_{e}$ from the protoneutron
star, and hence, their emission energy spectra.
An interesting issue is whether strong magnetic
fields in supernovae could leave detectable imprints on the neutrino
energy spectra. 

\section*{Acknowledgements}

HD is very grateful to the hospitality of Tony Mezzacappa, George M. Fuller,
and the Institute for Nuclear Theory during the workshop.
He also wants to thank Arkady Vainshtein for helpful discussions. This work
was supported in part by DOE grant DE-FG02-87ER40328.

\end{document}